\documentclass{aa}
\usepackage{graphicx}

\def\apj{ApJ}
\def\apjs{ApJS}
\def\aa{A\&A}
\def\aas{A\&AS}
\def\araa{ARA\&A}
\def\apss{Ap\&SS}

\def\aj{AJ}
\def\mn{MNRAS}
\def\nat{Nat}

\def\h92{H92$\alpha$\/\ }

\def\l0{{l$_0$\/\ }}

\def\asecc{$^{\prime\prime}$}
\def\aminn{$^{\prime}$}

\def\degg{$^{\circ}$}
\def\msun{$M_{\odot}$\/\ }
\def\msunyr{$M_{\odot}$ yr$^{-1}$\/\ }
\def\msunnyr{$M_{\odot}$ yr$^{-1}$}
\def\lsun{$L_{\odot}$\/\ }
\def\lsunn{$L_{\odot}$}
\def\msunn{$M_{\odot}$}

\def\kms{km s$^{-1}$}
\def\sp{$\alpha^{850}_{1.4}$\/\ }

\begin{document}
\title{Search for sub-mm, mm and radio continuum emission from Extremely Red Objects}
\author{Niruj R. Mohan\inst{1,2} \and A. Cimatti\inst{3} \and H.J.A. R\"ottgering\inst{4} \and P. Andreani\inst{5,6} \and P. Severgnini\inst{7} \and \\ R.P.J. Tilanus\inst{8} \and C.L. Carilli\inst{9} \and S.A. Stanford\inst{10}}
\institute{Raman Research Institute, C.V. Raman Avenue, Sadashivanagar, Bangalore 560080, India \and Joint Astronomy Program, Department of Physics, Indian Institute of Science, Bangalore 560012, India \and Osservatorio Astrofisico di Arcetri, Largo Fermi 5, 50125 Firenze, Italy \and Sterrewacht Leiden, Sterrewacht, Postbus 9513, Leiden 2300 RA, The Netherlands \and Max-Planck Institut f\"ur Extraterrestrische Physik, Postfach 1312, D-85741 Garching bei Muenchen, Germany \and Osservatorio Astronomico di Padova, vicolo dell' Osservatorio 5, 35122 Padova, Italy \and Dip. Astronomia, Universita' di Firenze, Largo E. Fermi 5, 50125 Firenze, Italy \and Joint Astronomy Centre, 660 N. Aohoku Place, Hilo, HI 96720, USA \and NRAO, P.O. Box 0, Socorro NM, 87801, USA \and Institute of Geophysics and Planetary Physics, Lawrence Livermore National Laboratory, L-413, P.O. Box 808, Livermore, CA 94550, USA}
\offprints{Niruj R. Mohan}
\mail{niruj@rri.res.in}
\date{Recieved / Accepted  }
\titlerunning{EROs : radio and sub-mm continuum from EROs}
\authorrunning{N. R. Mohan et al}

\abstract{ 
We present the results of sub-mm, mm (850$\mu$m, 450 $\mu$m and
1250$\mu$m) and radio (1.4 and 4.8 GHz) continuum observations of a
sample of 27 $K$-selected Extremely Red Objects, or EROs, (14 of which
form a complete sample with $K<20$ and $I-K>5$) aimed at detecting
dusty starbursts, deriving the fraction of UltraLuminous Infrared 
Galaxies (ULIGs) in ERO samples, and constraining their redshifts using
the radio-FIR correlation. One ERO was tentatively detected at 
1250$\mu$m and two were detected at 1.4 GHz, one of which has a less
secure identification as an ERO counterpart.
Limits on their redshifts and their star forming properties are derived
and discussed. We stacked the observations of the undetected objects
at 850$\mu$m, 1250$\mu$m and 4.8 GHz in order to search for possible
statistical emission from the ERO population as a whole, but no
significant detections were derived either for the whole sample or
as a function of the average NIR colours. These results strongly suggest
that the dominant population of EROs with $K<20$ is not comprised of
ULIGs like HR 10, but is probably made of radio-quiet ellipticals
and weaker starburst galaxies with L$\,<\,$10$^{12}$ \lsun and SFR$\,<\,$100 \msunyr.
\keywords{Galaxies: elliptical and lenticular, cD -- Galaxies: general -- Galaxies: high-redshift -- Infrared: galaxies -- Radio continuum: galaxies -- Submillimeter} }
\maketitle

\section{Introduction}

The existence of a population of extragalactic objects with extremely red 
infrared-optical colours has been known for a number of years (for a recent review, see
Cimatti 2000\nocite{cimdf00}). These objects 
were initially discovered mainly in near-infrared surveys of blind fields and quasar 
fields and are very faint or invisible in the optical bands (Elston et al. 1988\nocite{err88}; 
McCarthy et al. 1992\nocite{mcpw92}; Hu \& Ridgway 1994\nocite{hr94}). These extremely 
red objects (EROs) are defined as those which have $R-K$ colours $\geq$ 5 and these 
tend to have $K$ magnitudes fainter than $\sim$18. Since the time of their discovery, 
the nature of this population
has remained a puzzle. Based on their NIR and optical photometric data, two broad
classes of models are consistent with the observed red colours : (a) high redshift
starbursts, red because of severe dust extinction. From the required extinction and
simple dust models, these galaxies could be normal starburst galaxies or even high-$z$
counterparts of local Ultra Luminous Infrared Galaxies (ULIGs) and (b) old passively 
evolving ellipticals at redshifts greater than about one. The red colours of the 
EROs would then be explainable by a large  $K$-correction and an absence of ongoing 
star formation. If the EROs belong to the former class, then they would be dominant 
sites of star formation and would be important in determining the star formation 
history of the universe (Cimatti et al. 1998a\nocite{nat98}). On the other hand, 
if they belong to the latter class, then the volume density of these objects
as a function of redshift would pose strong constraints on the models for the formation 
of elliptical galaxies, which range from monolithic collapse to dark matter 
dominated hierarchical structure formation scenarios (Daddi et al. 2000b\nocite{dcr00} 
and references therein).

Over the last few years, it has been possible to study in detail a handful of EROs
which are bright enough to yield reliable spectra and have multi-wavelength continuum
data. These studies were able to determine the nature of these EROs and there are 
now examples known for both starbursts (Cimatti et al. 1998a\nocite{nat98}; Cimatti 
et al. 1999\nocite{cim99}; Smail et al. 1999a\nocite{smail99a}; Smail et al. 
1999b\nocite{smail99b}; Gear et al. 2001\nocite{gear}) as well as for old ellipticals 
(Spinrad et al. 1997\nocite{sp97}; Cimatti et al. 1999\nocite{cim99}; Soifer et al. 
1999\nocite{soi99}; Liu et al. 2000\nocite{liu00}) among the ERO population. 
Studies indicate though, that not more than $\sim$30 \% of EROs are 
starbursts (see Sect. 5 for details). Recently, independent wide-field surveys 
for EROs have been conducted (Daddi et al. 2000a\nocite{daddi00}; McCarthy et 
al. 2000\nocite{mccar00}; Thompson et al. 1999\nocite{thomp99}) which have shown 
that these objects are strongly clustered in the sky (see also Chapman et al. 
2000\nocite{chapman00} and Yan et al. 2000\nocite{yan00}). The surface density 
of these EROs after correcting for clustering, assuming that these are passively 
evolving ellipticals, has been shown to be consistent with pure luminosity 
evolution with a formation redshift greater than 2.5 (Daddi et al. 
2000b\nocite{dcr00}). 

HR 10 is one of the reddest EROs and is quite bright in the NIR ($I-K$=6, 
$K$=18.42, Graham and Dey 1996\nocite{gd96}). \cite{nat98} 
detected 850 $\mu$m and 1250 $\mu$m emission from this object (see also Dey 
et al. 1999\nocite{dey99}). They derived a star formation rate of several hundred \msunyr 
and an FIR luminosity in excess of of 10$^{12}$ \lsunn, thus showing that HR 10
is an ULIG at its redshift of 1.44 (Graham and Dey 1996\nocite{gd96}). At the 
time of the observations presented in this paper, HR 10 was the only ERO with 
detected submm emission and also the only ERO with a known redshift. Hence 
it was thought possible that a majority of EROs
would be similar to HR 10 and would have observable submm and mm emission 
(Cimatti et al. 1998b\nocite{cim98b}, Andreani et al. 1999\nocite{and99}). Therefore 
we began a search for mm and submm continuum emission from other EROs with an aim to
detect extreme starburst galaxies. Further, in order to constrain the redshifts 
of these objects and also to understand their nature, we also decided to search 
for radio continuum emission from these EROs. 

The radio and the FIR luminosities of nearby galaxies ($z<\,$0.4) which are 
dominated by star formation are known to be highly 
correlated (see Condon 1992\nocite{condon}
and references therein), whereas E and S0 galaxies are known to be more 
radio bright than a star-forming galaxy of similar FIR luminosity (Walsh et al. 
1989\nocite{wkwk89}). The radio (mainly synchrotron and some amount of
free-free) emission and the FIR (due to dust) emission have different spectral 
indices. Hence, assuming the local radio-FIR correlation holds at high redshift,
the observed ratio of radio to FIR emission strengths can be used to determine
the redshift of a star-forming galaxy (the redshift determination method 
and the associated error estimation are developed in Carilli and 
Yun (1999, 2000)\nocite{cy99,cy00} and Blain 1999\nocite{blain99}).  

So, if on the one hand, the ERO population 
consists primarily of starbursts, this method could be used to determine the nature of 
EROs and also to constrain the redshifts of these objects. If, on the other hand, 
the EROs are ellipticals, their IR colours would be used to constrain their
properties. Since most EROs are too faint to obtain redshifts even with 10m-class 
telescopes or even to obtain accurate photometry over the entire optical-IR 
range, such complementary diagnostics become important in understanding these 
objects.

We first describe the ERO sample, the observations and their results 
in Sects. 2 and 3. In Sect. 4, we derive statistical properties of the 
sample and estimate the fraction of starbursts and ellipticals with radio 
emission in Sect. 5. Those EROs with radio or mm detections are discussed 
in detail and their properties are derived in Sect. 6. The cosmology 
adopted in this paper is a flat universe with H$_{\rm 0}$=70 \kms~Mpc$^{-1}$ and 
all results are calculated for both $\Omega_\Lambda$=0.7 and $\Omega_\Lambda$=0.0. 
The spectral index is assumed to be $-$0.7 throughout this paper for calculating 
the $K$-correction for the radio emission.

\section{Sample and Observational details}

A sample of 27 EROs was selected from the literature and also from other 
samples being studied by the authors. These objects were identified from 
a variety of observations : deep NIR surveys of random fields, galaxy cluster 
fields and quasar fields (see references in Table 1). A $R-K$ or an $I-K$ 
lower cut-off of 5 was used to select these EROs. Given the heterogenous
nature of the fields from which these objects were taken, the sample as a 
whole is not complete or uniform. However, a subset of the objects observed, 
the EES sub-sample (Elston et al. 2001\nocite{eeal01}) does form a {\it 
complete} sample of 14 EROs, selected such that $K<20$ and $I-K>5$, and are listed 
separately in Table 1. Hereafter, the term ERO will imply all 27 objects listed 
in the table and the EES sample will include only the 14 objects forming a complete 
sample. The average $R-K$ and $I-K$ colours of the EES sample and the 
rest of the EROs are statistically indistinguishable. A sub-sample of 
the EROs selected were observed at 1250 $\mu$m using the IRAM 30 m 
telescope (15 objects) and at 850 $\mu$m (21 objects) and 450 $\mu$m 
(5 objects) using the SCUBA at the JCMT. Some of these objects have 
also been observed either at 4.8 GHz or at 1.4 GHz (11 and 7 objects 
respectively) using the VLA. Due to various telescope scheduling constraints, 
the same objects could not be observed at all wavebands. 

\subsection{NIR observations of the EES complete sub-sample}

Elston et al. (2001\nocite{eeal01}) have completed a $BRIzJK$ field 
survey (hereafter, EES survey) covering $\sim$100 arcmin$^2$ over
four areas of the sky at high galactic latitudes, down to $K$$\sim$21.5. These
observations were carried out at the 4~m telescope of the Kitt Peak National
Observatory.  Optical imaging was obtained with the PFCCD/T2KB, which gives
0.48$\arcsec$ pixels over a 16$\arcmin$ field. The IR imaging was obtained
with IRIM, in which a NICMOS3 HgCdTe array provides 0.6$\arcsec$ pixels over
a $\sim$2.5$\arcmin$ field.  Details of the observations and data analysis are
given in Elston et al. (2001). The $BRIzJK$ survey images were co-aligned and 
convolved to the same effective PSF of FWHM$\sim$1.5$\arcsec$. Calibrations of 
the optical and IR images onto the Landolt and CIT systems were obtained using
observations of Landolt and UKIRT standard stars, respectively.

A catalogue of objects within each of the four fields of the EES survey
was obtained from the $K$ images using SExtractor (Bertin \& Arnout 1996\nocite{ba96}).  
Object detection was performed down to a level corresponding 
to 1.5 $\sigma$ above the sky level, with a minimum object size of 1.2
square arcseconds. Photometry was obtained through 3.3$\arcsec$ diameter apertures
using this catalog on the $BRIzJK$ images; the 4 $\sigma$ limit in the $K$ band
is 21.4.  Details of these catalogues are available in Elston et al. 
(2001)\nocite{eeal01}.

\subsection{SCUBA observations at the JCMT}

A sub-sample was observed with Submillimeter Common-User Bolometer Array 
(SCUBA) instrument on the James Clerk Maxwell
Telescope (JCMT) at 450 $\mu$m and 850 $\mu$m
in the standard point-source photometry mode during different
observing runs from 1998 to 2000. The typical opacity $\tau$~was 
in the range 0.2--0.5 and 2--4 at 850 $\mu$m and 
450 $\mu$m respectively. The on-source integration times varied between 
900 and 3600 seconds. The data reduction was performed using the 
Starlink SURF software (Jenness \& Lightfoot 1998\nocite{jl98}). For each integration, 
the measurements in the reference beam were subtracted from those in 
the signal beam, rejecting obvious spikes. Flat-field corrections were 
applied to each observation which were subsequently corrected for 
atmospheric opacity. The residual sky background emission was then removed
using the median of the outputs of the different rings of the bolometer 
as a background estimate.
The flux calibration of the data was performed mainly using a primary calibrator 
(Uranus or Mars), thus yielding a 10\% accuracy for the
flux density scale. A poorer calibration accuracy (20\%) was obtained 
for EES Lynx~2  and SA57~2--3 for which the secondary calibrators 
CRL618 and HL Tau were used respectively. The individual reduced and calibrated 
observations were concatenated for each source thus obtaining a final co-added 
data set. 

\subsection{IRAM 30m telescope observations}

The 1.25 mm data reported here were taken with the MPIfR 37-channel
and 19-channel bolometers \cite{kreysa98} at the focus of the 
IRAM 30m antenna (Pico Veleta, Spain) during observing runs in 
March 1998 and December 1998 respectively.
The filter set used combined with atmospheric transmission produces an effective
wavelength of about 1.25 mm; the beam size is $11 ^{\prime \prime}$
(FWHM) and the chop throw was set at $50^{\prime \prime}$ and $30^{\prime
\prime}$, during the first and second run, respectively. The expected
average sensitivity for each channel, limited principally by atmospheric noise,
was 60 mJy/$\sqrt{BW}$, where BW is the bandwidth used, in Hz. 
However, the 37-channel bolometer observations 
were noisier than expected and therefore the 19-channel bolometer was 
used for the second observing run. The effect of sky noise was
reduced substantially by exploiting the correlation between signals 
from the different channels using the standard three beam (beam-switching + 
nodding) technique, resulting in an average rms 1$\sigma$ value, after 
6000 seconds of integration, of 0.57 mJy and 0.4 mJy for 
the 37-channel and 19-channel data respectively. The typical on-source 
integration time was 6000 seconds, distributed over two to three 
nights. The atmospheric transmission 
was monitored by making frequent sky-dips. The average zenith 
opacity was 0.13 during the observations in March, with a maximum of 0.2, and 
the average opacity was $\sim$0.3 during the December observing run.
Absolute flux calibration was performed using Uranus as the 
primary calibrator and using Mars and quasars from the IRAM 
pointing list as secondary calibrators. The different calibration
measurements were consistent at a level of 5 \% for both planets. Including 
the uncertainty in the planet temperatures, the average flux 
calibration uncertainty was estimated to be 10 \%. Pointing was 
checked every hour and the average accuracy achieved was 
better than $3 ^{\prime \prime}$.

The data were reduced assuming that the target sources are unresolved,
i.e. the source sizes at mm wavelengths are smaller than the size of
the central channel. The remaining 36 and 18 channels (excluding one which
suffered a large electronic loss) were then used to derive a
low-noise sky estimate. The average value of the sky brightness, computed using
these outer 35 and 17 channels respectively, was subtracted from the signal in 
the central channel to derive the final flux density estimates.

\subsection{VLA observations}

The Very Large Array (VLA) was used to observe a total of 16 EROs distributed in six 
different fields. 
4.8 GHz radio continuum emission was searched for in eleven of these objects 
in the D configuration at 4.8 GHz, chosen as a compromise between 
optimising point source sensitivity using a low resolution array and minimizing 
confusion through the usage of a high resolution array. 
The correlator was used in the continuum mode and data were acquired in 4 IF bands, 
each of 50 MHz bandwidth. The flux calibration and initial phase calibration 
was done using standard techniques described in \cite{bible98} and the data 
reduction was done using standard algorithms in the software AIPS. Self-calibration
of only the visibility phases was done for a few fields using background sources 
in the primary beam. The rms noise in the final images is within 20\% of the 
expected thermal noise. 

Seven EROs distributed in two fields were observed using the VLA in the B configuration 
at 1.4 GHz. The frequency of observation was chosen based on the expected steep spectral 
index of the continuum emission of EROs (typically $-$0.7) and poor sensitivity of the 
VLA at $\nu$$\,<\,$1 GHz. Optimising between confusion and the slightly 
extended nature of EROs
($\geq$2\asecc) led to the B configuration being used (the number of sources within a  
synthesised beam area in this configuration with flux densities higher than  1$\sigma$ for 
8 hours of integration is 0.015, using the relation given in Langston et al. 
1990\nocite{lang90}). These observations were made in the line mode using 4 
IFs, each with a bandwidth of 25 MHz and 7 available channels (the 50 MHz 
bandwidth system was not used at this frequency as much higher closure errors 
are expected: Owen, private communication). The dataset were hanning smoothed 
following standard calibration procedures and were imaged using multi-band synthesis 
(in order to reduce the effect of bandwidth smearing far from the phase centre).
The tangent-plane approximation was used to correct for the effects of the array 
non-coplanarity and strong sources within two primary beams of the phase centre were 
isolated by defining multiple fields and simultaneously deconvolved using the CLEAN 
algorithm. Self-calibration using sources in the primary beam was done only for the 
visibility phases and the data were flagged based on excessive closure errors. An rms 
within 20\% of the theoretical noise was achieved for the EES-Pisces field but the 
noise in the EES-Cetus field was substantially higher due to the presence of strong 
sources near the phase centre. Further details are given in Table 2. 

\section{Results}

Out of the seven sources observed at 1250 $\mu$m, one source, SA57-1, was 
marginally detected at the 3$\sigma$ level. The measured flux densities at 
the NIR positions of the EROs along with the 1$\sigma$ errors are listed 
in Table 3 for the 450 $\mu$m, 850 $\mu$m and the 1250 $\mu$m data. As can 
be seen, none of the other observed sources were detected at these wavelengths.

\begin{figure}
\resizebox{\hsize}{!}{\includegraphics{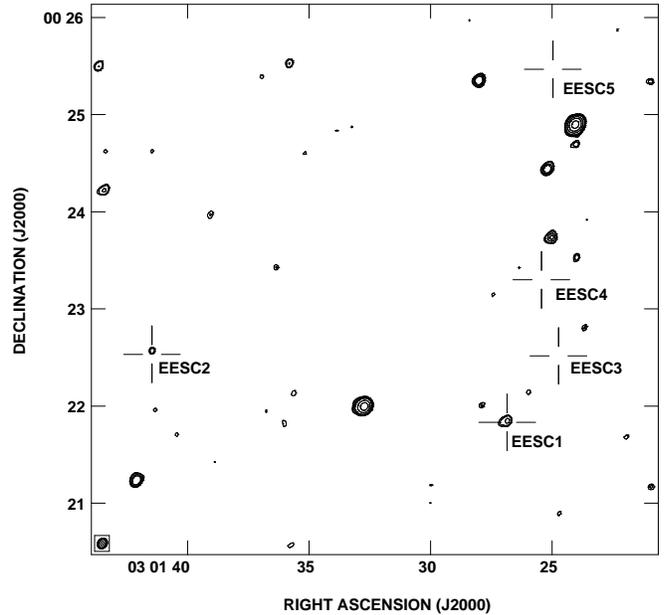}}
\caption{Contour plot of the 1.4 GHz continuum image of the EES-Cetus 
field. The centres of the stars mark the NIR positions of the EROs. The two detections, 
EESC1 and EESC2 can be seen. The CLEAN beam is 6.5\asecc$\times$5.3\asecc 
and is shown in the bottom left corner of the figure. The rms is 75 $\mu$Jy. 
The contour levels are at 3.3$\sigma$, 4$\sigma$, 6$\sigma$, 8$\sigma$, 
17$\sigma$, 32$\sigma$ \& 63$\sigma$.}
\label{fig1}
\end{figure}

We have detected 1.4 GHz radio continuum emission from the NIR positions of 
two EROs : EES-Cetus 
1 and EES-Cetus 2 (hereafter, EESC1 and EESC2). None of the other sources 
observed at either 4.8 GHz or 1.4 GHz were detected and the corresponding 
measurements at the expected NIR positions and the 1$\sigma$ errors are 
given in Table 3. The 1.4 GHz image of the EES-Cetus field with the two 
detections is shown in Fig. 1. Though there appears to be some extended emission
in the image of EESC1, given the signal-to-noise ratios of the two detections, the
two sources are essentially unresolved. We estimate the likelihood of these two 
radio sources being the counterparts of the NIR-detected EROs as follows : 
The positional error in the radio positions is calculated for
the two sources using the relations given in Rieu (1969)\nocite{rieu}, assuming
the error in the NIR position to be 0.5\asecc, which is 
the 1$\sigma$ error between the radio and the optical reference frames (Russell
et al. 1990)\nocite{r90}. The value of the likelihood ratio (LR) as defined by
de Ruiter et al. (1977)\nocite{rui77} is calculated to be 1276 for EESC1 and 43 for 
EESC2, using the relation in Langston et al. (1990)\nocite{lang90} to estimate the
surface density of objects with S$_{\rm 1.4~GHz}>\,\,$0.1 mJy. This implies that
the {\it a posteriori} probabilities of the radio detections being actual counterparts
are 99.92 \% and 97.73 \% for EESC1 and EESC2 respectively (assuming that the {\it a 
priori} probability that the ERO does have a radio counterpart is 50 \%). Hence
the radio counterpart to EESC1 seems real but the identification of the counterpart of
EESC2 is slightly less secure. 

\section{Statistical properties of the sample}

Though almost the entire sample is undetected at all observed wavelengths,
the flux densities measured at the IR positions of the sources can be used 
to compute the statistical
flux densities of the sample and hence derive stricter upper limits.
This weighted average flux density at 4.8 GHz is 3.3 $\pm$ 4.6 $\mu$Jy for a 
sample of 10 objects observed (the value for the 1.4 GHz data is $-$15 $\pm$ 
12 $\mu$Jy, and is far less significant in terms of the implied limits on the
star formation rate). The statistical flux density at 850 $\mu$m 
for the entire sample of 21 EROs is 1.0 $\pm$ 0.4 mJy and the corresponding
number for the 1250 $\mu$m measurements is 0.18 $\pm$ 0.12 mJy for 15 sources. 
Clearly, the 1250 $\mu$m and the 4.8 GHz data do not yield a statistical detection
of the ERO population whereas the 850 $\mu$m data yields a marginal 2.5$\sigma$
statistical detection. A similar exercise was carried out for the EES 
sample as well and the average flux densities at 4.8 GHz, 1250 $\mu$m 
and 850 $\mu$m are 3.5 $\pm$ 4.8 $\mu$Jy, 0.18 $\pm$ 0.13 mJy and 0.3 $\pm$ 0.6 mJy 
for nine, seven and fourteen objects respectively; there is no statistical detection
for this complete sample either. We have quoted the values for the 4.8 GHz and 
the 1250 $\mu$m data for the EES sample though not all sources have been
observed, as the observed sources probably form a random sub-sample of the complete 
sample since they were selected based on telescope scheduling constraints. 
The 2.5$\sigma$ result for the average 850 $\mu$m
flux density for the entire sample can be traced solely to the non-EES sample of 
7 galaxies, whose mean signal to noise ratio is 1$\sigma$. Therefore it is probable
that some of these EROs not in the EES sample might be detectable at 850 $\mu$m if observed
further. In this paper, we will take the 3$\sigma$ upper limit of 1.2 mJy for the
average flux density at this wavelength for the entire ERO sample.

The sample was also divided into two sub-samples using an $I-K$ colour cut-off 
(various cut-off values ranging from 5.3 to 5.9 were tried) and it was seen that 
there is no statistically significant difference in the average flux densities 
between the two sub-samples. This result holds for the EES sample as well.
If the observed population is divided into ellipticals and starbursts 
using the $I-K$ versus $J-K$ diagram of Pozzetti \& Mannucci (2000)\nocite{pm00}, 
described in Sect. 5.1, the two categories of EROs do not differ in their
average flux densities either, within errors.

\section{Ellipticals or Starbursts ?}

Though attempts to distinguish between pure starbursts and elliptical
galaxies among the ERO population have met with unambiguous results only
for sources bright in IR and optical, or for sources with detectable 
mm/sub-mm emission, a few recent studies seem to show 
that probably $\geq$ 70\% of the EROs are old high redshift ellipticals. 
The evidence for this is based on both individual spectroscopic 
identifications of small samples of EROs (Cimatti et al. 1999\nocite{cim99}; Liu et al. 
2000\nocite{liu00}) and on high resolution images, using morphological
information (Stiavelli \& Treu 2001\nocite{stia}) and through fitting the 
de Vaucouleurs' law to radial profiles (Moriondo et al. 2000\nocite{mor00}).
Additionally, a recent wide-field survey of EROs by \cite{daddi00}
has shown that these objects are strongly clustered in the sky and 
this result has been confirmed in two other fields by McCarthy et 
al. (2000)\nocite{mccar00}. The strong clustering of the ERO population
is added evidence that the majority of EROs are indeed ellipticals, since
ellipticals are known to be more clustered than spirals, and also because of
the narrow range of redshift, $z$=1--2.5, allowed for extremely red 
ellipticals (Daddi et al. 2000b\nocite{dcr00}).

\subsection{Diagnostic techniques}

In this section, the inferences that can be drawn from the observed 
multi-frequency flux densities of the EROs using group properties of 
ellipticals and starbursts are investigated.

It is known that the $K$ magnitudes of radio-loud galaxies are correlated with 
their redshifts (Lilly \& Longair 1984\nocite{ll84}). It is also established 
that galaxies of other types have fainter $K$ magnitudes than radio-loud
galaxies at that redshift (van Breugel et al. 1999\nocite{breu99}, de 
Breuck 2000\nocite{carlos}), which can be used to set upper limits on the
redshift of EROs. Given the faintness of EROs in the $K$ band, such an exercise
constrains them to lie at $z<$5. 

If we consider the upper limit to the 1.4 GHz flux density of the undetected
ERO sample to be 0.1 mJy, (extrapolated from the 4.8 GHz upper limit. This excludes
the EROs observed at 1.4 GHz which have varying upper limits and excludes the two 
detections as well) then for $z<5$, the rest frame 1.4 GHz luminosity is less than 
2$\times$10$^{25}$ W/Hz for $\Omega_\Lambda$=0.7 (and less than 6$\times$10$^{24}$ 
W/Hz for $\Omega_\Lambda$=0). From the bi-modal 1.4 GHz luminosity distribution of 
the IRAS 2 Jy sample (fig. 15 of Yun et al. 2001\nocite{yrc01}), it is clear that
our sample cannot be differentiated into galaxies dominated by starbursts versus
those by AGNs, which reflects the fact that our radio data is not deep enough to 
detect weak starbursts. 

Pozzetti \& Mannucci (2000)\nocite{pm00} showed that old ellipticals and dusty
starbursts occupy distinctly different areas in the $I-K$ versus $J-K$ diagram
and derived the theoretical dividing line in this plane. Due to the faintness of
most EROs in the $K$ band, the error bars for the colours are too large for this
diagnostic to be used profitably. Additionally,
HR 10, the definitive example of a dusty starburst ERO (Andreani et al. 
2000\nocite{aclr00}; Dey et al. 1999\nocite{dey99}; Cimatti et al. 1998a\nocite{nat98}),
lies on the dividing line. Hence, more sensitive NIR photometry is needed in order
to use this method of classification. 

For our sample, based on the few IR-optical colours 
available, and the upper limits from the radio, mm and sub-mm observations, 
it is not possible to determine the nature of each of the EROs individually. 
Instead, we now discuss the statistical properties of EROs, for the dusty 
starburst galaxy and the old elliptical galaxy components, seperately. 

\subsection{Star formation properties of EROs}

The average 850 $\mu$m flux density limit can be used to derive 
upper bounds on the average star formation properties of the sample.
Assuming that all EROs in the sample are at a redshift of 1.5 
(i.e., roughly at the redshift of HR 10), 
we derive the average star formation rate to be less than 150 \msunyr 
(from the relation given in Carilli \& Yun 1999\nocite{cy99}, and assuming
the dust emissivity $\beta$=1.5)
and the average FIR luminosity to be less than 1.6$\times$10$^{12}$ \lsun 
($\Omega_\Lambda$=0). For a dust temperature of 20 K, the corresponding 
upper limit on the average dust mass is 6$\times$10$^8$ \msun and for 
50 K, the dust mass is less than 1.4$\times$10$^8$ \msunn. For $\Omega_\Lambda$=0.7,
SFR $<$ 380 \msunnyr, L$_{\rm FIR}<$ 4$\times$10$^{12}$ \lsun and M$_{\rm dust} <$ 2$\times$
10$^9$ \msun for T$_{\rm dust}$=20 K and $<$ 3.6$\times$10$^8$ \msun for T$_{\rm dust}$=50 K
(for a dust emissivity $\beta$=1.5 and mass absorption coefficient of 0.15 m$^2$ 
kg$^{-1}$ at 800 $\mu$m; Hughes et al. 1997\nocite{hdr97}).

However, we can assume that a fraction $x$ of the EROs are dusty ULIGs
which resemble HR 10 in their properties (L$_{\rm FIR}\sim$4$\times$10$^{12}$ 
\lsun at z$\sim$1.5) and the rest are ellipticals with no sub-mm and mm
emission. Then, from the estimated 5$\sigma$ average flux densities of the 
ERO sample at 850 $\mu$m and 1250 $\mu$m and the measured flux densities
of HR 10 (S$_{\rm 850~\mu m}$=5.5 $\pm$ 0.6 mJy, the weighted average of the 
values quoted in Cimatti et al. 1998a\nocite{nat98} and Dey et al. 
1999\nocite{dey99} and S$_{\rm 1250~\mu m}$=4.9 $\pm$ 0.7 mJy, Cimatti et 
al. 1998a\nocite{nat98}), the value of $x$ can be computed. Such an exercise
gives $x<$ (36 $\pm$ 11) \% and $x<$ (12 $\pm$ 4) \% for the 850 $\mu$m 
and 1250 $\mu$m data respectively. This estimate, though approximate, is
consistent with other independent estimates of the ULIG fraction in EROs 
: $\leq$30 \% (Cimatti et al. 1999\nocite{cim99}; Moriondo et al. 
2000\nocite{mor00}).  

If we assume, as a conservative estimate, that the starburst fraction in the ERO
population is as much as 30 \%, then from the surface density of EROs with $R-K_s\geq5$ 
and $K_s\leq19.2$ estimated by \cite{daddi00}, the surface density of starbursts
would be less than 725 $\pm$ 33 objects deg$^{-2}$. \cite{smail99a} calculated
that the surface density of SCUBA sources with a 850 $\mu$m flux density greater
than 0.5 mJy (which corresponds to the cut-off estimated in order to fully explain the
Far Infrared Background or FIRB) is 17000 objects deg$^{-2}$ (for S$_{\rm 850~\mu m}\geq2$ 
mJy which would explain
half the observed FIRB, the surface density is 3700 objects deg$^{-2}$). Hence the
maximum overlap between $R-K_s\geq5$ and $K\geq19.2$ EROs which are starbursts, and the high
redshift star forming SCUBA sources is 4 \% (for a 0.5 mJy cut-off for S$_{\rm 850~\mu m}$; it 
is less than 20 \% for a 2 mJy cut-off).

Given the lack of detectable sub-mm emission from the sample of objects, 
and the estimate of the corresponding fraction of HR 10-like ULIGs, the 
present study clearly shows that the ULIGs like HR 10 are rather unique 
objects among EROs and hence dusty strong starbursts are not the dominant component
of this population. 

\subsection{EROs as elliptical galaxies}

It can be seen from the upper limits to the radio luminosities of the EROs 
(derived in section 5.1) that there are no radio-loud ellipticals in the sample. 
Hence these EROs must either be centre-brightened radio galaxies (or FR I; see
Fanaroff \& Riley 1974\nocite{fr74} and Ledlow \& Owen 1996\nocite{lo96} for
definitions. The optical luminosities have been derived assuming a median 
redshift of 1.5 and m$_R$ derived from Table 1) or radio-quiet ellipticals.
There are two detections at 1.4 GHz, and the 4.8 GHz upper limits of 11 EROs 
scaled to 1.4 GHz are $\sim$0.1 mJy. Assuming that $\geq$70\% of these 13 
EROs are elliptical galaxies, the detection rate of ellipticals for a 0.1 
mJy cut-off at 1.4 GHz is calculated to be $\leq$22 $\pm$ 16 \%. If the three
EES-Cetus sources undetected in the radio are also included, the detection rate
becomes $\leq$27$\pm$ 19 \%. For a redshift range of 1--3, 0.1 
mJy at 1.4 GHz corresponds to a 1.4 GHz rest-frame radio luminosity 
of 2$\times$10$^{23}$-2$\times$10$^{24}$ W/Hz ($\Omega_\Lambda$=0). 
Also, the rest-frame $R$ band magnitude for the sample is between $-$18 
and $-$25.5 (assuming a $K$-correction $K$($z$)=1.122$z$ for elliptical 
galaxies, see Ledlow \& Owen 1996\nocite{lo96}). It should be noted that for
the same set of parameters, using the data published in Gavazzi \& Boselli 
(1999)\nocite{gb99}, Ho (1999)\nocite{ho99}, Ledlow \& Owen (1996)\nocite{lo96} 
and Auriemma et al. (1977)\nocite{auri77}, we derive the detection rate to be 
1\% to $<$3\% for low redshift ellipticals.  

\section{Continuum detections -- EESC1, EESC2 and SA 57-1}

The two EROs, EESC1 and EESC2, with 1.4 GHz continuum detections (though the radio
identification is tentative for EESC2) have only upper 
limits to their sub-mm flux densities. The optical-IR SED of these objects cannot 
be used to classify them as either elliptical or starforming galaxies unambiguously. 
Further, the upper limit to the 850 $\mu$m-1.4 GHz spectral index \sp is +0.47 and 
+0.56 for EESC1 and EESC2 respectively. 

The ratio of radio to sub-mm flux density of a pure starburst decreases 
with increasing redshift, 
i.e., for a given sub-mm flux density, the $z$=0 galaxies have the maximum 
radio flux density. Hence for a given upper limit to the sub-mm flux density for
a galaxy at an arbitrary redshift, the $z$=0 radio-FIR correlation will yield the
maximum possible radio flux density. Therefore a sufficiently radio-bright AGN 
might possibly have a radio continuum strength 
higher than this value and will be easily classified as a radio-loud AGN. Since that
is not so for either of the two radio detections, they cannot be unambiguously
classified as either an old elliptical or a dusty starburst.
The properties of these galaxies are derived below assuming either of the two
possibilities -- that they are pure starbursts, or they are old ellipticals.

\subsection{EESC1 and EESC2 : 1.4 GHz detection}

If EESC1 and EESC2 are assumed to be dominated by star formation, 
then from the derived upper limits to the spectral index \sp for EESC1 
and EESC2, we can derive the upper limits to the redshifts of these 
objects. \cite{cy00} have tabulated the values of redshift for a given 
value of the spectral index \sp and have also tabulated the values 
for the $\pm$1$\sigma$ curves of redshift versus \sp. Using the 
derived value of $\alpha$+$\Delta\alpha$, where $\Delta\alpha$ is 
the 1$\sigma$ error on the value of the spectral index $\alpha$, 
and the z$^-$ curve of \cite{cy00}, updated in Carilli (private 
communication)\footnote{The updated values can be found at the URL: 
http://www.aoc.nrao.edu/$\sim$ccarilli/alphaz.shtml.}, we estimate 
the upper limits to the redshifts (see Carilli \& Yun 2000\nocite{cy00} 
for details).

The limiting redshifts of EESC1 and EESC2, derived as described above, 
are $z<~$1.5 and $z<~$2.0 respectively. Their radio continuum flux 
densities, if attributed solely to star formation, imply a star formation 
rate of about 1100 \msunyr and 1600 \msunyr respectively for the 
maximum redshifts derived above (these are for $\Omega_\Lambda$=0.7. 
The values are about 700 \msunyr and 1000 \msunyr respectively, 
for $\Omega_\Lambda$=0), and less than this value for lower values of 
redshift (from the relation given in Carilli \& Yun 1999\nocite{cy99}). 
Also, from the observed radio flux density, the FIR luminosity is 
calculated to be less than $\sim$10$^{13}$ \lsun, for $z<z_{\rm max}$, 
for the two galaxies.

If the two galaxies are assumed to be old ellipticals instead, then
using the pure luminosity evolution models, the extreme $R-K$ colours 
of the two galaxies ($R-K>$6.1) imply that they lie at redshifts greater 
than 1.3 (Daddi et al. 2000b)\nocite{dcr00}. For $z$ between 1.3
and 5, the 1.4 GHz rest-frame luminosities of EESC1 and EESC2
correspond to the luminosity range of FR I galaxies or radio-quiet 
ellipticals.

\subsection{SA57-1 : 1.25 mm IRAM detection ?}

SA57-1 is detected at 1.25 mm with the IRAM 30m telescope 
at the 3$\sigma$ level: S$_{\rm 1.25~mm}$ =1.45$\pm$0.45 mJy. 
Assuming that this detection is real, it is not obvious whether 
this source is a pure starburst or whether it also harbours
an AGN. From the 1.25 mm 
measurement and the upper limit to the radio emission, a lower
limit of 0.9 can be derived for the redshift of the object (using
the $z^+$ curve in Carilli \& Yun 2000)\nocite{cy00}, as described 
in section 5.1. This is also the lower limit to $z$ if SA57-1 
is an elliptical galaxy. Using the $K-z$ relation for radio-loud 
galaxies, the derived upper limit to the redshift is 3.5. If 
this object were a pure starburst, the implied SFR is between 
45--800 \msunyr for $\Omega_\Lambda$=0 (and 70--1300 \msunyr 
for $\Omega_\Lambda$=0.7). The corresponding L$_{\rm FIR}$ is 
greater than 4.5$\times$10$^{10}$ \lsun. 

\section{Conclusions}

Motivated by the discovery of HR 10 and its star formation properties, a sample
of EROs was observed in order to detect radio, mm and sub-mm continuum emission
and constrain the redshifts and star formation rates of ULIGs in the sample. 
One ERO was detected at 1.4 GHz and a possible radio counterpart was identified
for another ERO at the same frequency. A third was tentatively identified at 1250 $\mu$m.
Their redshifts and star forming properties were constrained using their radio-FIR 
spectral index but their nature could not be unambigously determined. Since the 
sources are faint, standard techniques to classify the sources in the ERO 
sample individually as ellipticals or starburst are inadequate. Weighted average 
flux densities were computed for the sample using measurements at the IR positions of 
the EROs and these values are 1.0 $\pm$ 0.4 mJy at 850 $\mu$m, 0.18 $\pm$ 0.12 mJy at
1250 $\mu$m and 3.3 $\pm$ 4.6 $\mu$Jy at 4.8 GHz. We find no difference within 
errors in the weighted average values between EROs with NIR colours above and 
below an assumed $I-K$ colour cut-off. If the sample is divided 
into ellipticals and starbursts based on the NIR two-colour diagnostic diagram, no 
differences in their average flux densities are seen between these groups. 
From the lack of detection of sub-mm emission from any of the EROs in our sample, it
is now clear that dusty strong starbursts, or high redshift ULIGs, are not the 
dominant component of this population. If it is assumed that such an ULIG 
population would resemble HR 10 in their SED properties, then such galaxies cannot 
constitute more than about 35 \% of the EROs. From the observed source counts of
SCUBA sources and the ERO surface density, we suggest that EROs contribute negligibly
to both high redshift star formation as well as to the FIRB. On the other hand, if
all EROs have similar star formation properties, the average dust mass is calculated to 
be less than 2 $\times$ 10$^9$ \msun (for $\Omega_\Lambda$=0.7 and T$_{\rm dust}$=20 K) and 
the average FIR luminosity, less than 4 $\times$ 10$^{12}$ \lsunn. If not more than 
a third of the EROs in the sample are assumed to be starbursts and the rest are assumed 
to be ellipticals at a median redshift of 1.5, we calculate the detection rate of 
ellipticals for a 0.1 mJy cut-off at 1.4 GHz to be less than 22 
$\pm$ 16 \%. The corresponding number estimated in the local universe is $\leq$3 \%. 

Therefore it is clear that the dominant population of EROs are not ULIGs similar
to HR 10 but are probably old ellipticals and weaker starbursts. The determination 
of the properties and redshifts of the elliptical galaxy component of the ERO 
population is extremely important for constraining structure formation models 
(Daddi et al. 2000b\nocite{dcr00}).
Towards this end, the upper limits to the radio and the sub-mm flux densities 
derived in this study should be used to plan future observations to detect these 
objects at these wavebands.

\begin{acknowledgements}
The JCMT is operated by the Joint Astronomy Centre, on behalf 
of the UK Particle Physics and Astronomy Research Council,
the Netherlands Organization for Scientific Research (NWO)
and the Canadian research council. The VLA is a facility of 
the National Radio Astronomy Observatory (NRAO), which is 
operated by Associated Universities, Inc. under a cooperative 
agreement with the National Science Foundation. This work was 
based on observations carried out with the IRAM 30 m telescope. 
IRAM is supported by INSU/CNRS (France), MPG (Germany) and IGN (Spain).
This research has made use of NASA's Astrophysics Data System 
Abstract Service. We thank Wolfgang Tschager for help with the IRAM
observations.
Part of this work was done by NRM during his visit at Leiden Observatory 
and he thanks the Netherlands Research School for Astronomy (NOVA),
Leids Kerkhoven-Bosscha Fonds and Leiden observatory for financial 
support and the staff of Leiden Observatory for their hospitality and
stimulating discussions. NRM also acknowledges useful discussions with 
Frazer Owen and K.S. Dwarakanath about data analysis and K.R. Anantharamaiah
for encouragement in moving on to higher redshift realms.

\end{acknowledgements}

{}

\begin{table*}
\begin{center}
\caption[]{NIR and optical broadband magnitudes of EROs}
\begin{tabular}{lccccccc}
\hline
\multicolumn{1}{c}{Galaxy}& \multicolumn{1}{c}{RA} & \multicolumn{1}{c}{Dec} &\multicolumn{1}{c}{$K^a$} &\multicolumn{1}{c}{$J-K$} &\multicolumn{1}{c}{$I-K$} &\multicolumn{1}{c}{$R-K$} & Ref \\
\multicolumn{1}{c}{}& \multicolumn{1}{c}{(J2000.0)} & \multicolumn{1}{c}{(J2000.0)} &\multicolumn{1}{c}{(mag)} &\multicolumn{1}{c}{(mag)} &\multicolumn{1}{c}{(mag)} &\multicolumn{1}{c}{(mag)} & \\
\hline
\hline
\multicolumn{8}{c}{\bf EES Sample}\\
\hline
EES-Lynx 1  & 08$^h$48$^m$30$^s$.82 &+44\degg52\aminn51\asecc& 19.8 & 2.4 & 5.4    & $>$6.2 & 1 \\
EES-Lynx 2  & 08$^h$48$^m$43$^s$.60 &+44\degg53\aminn43\asecc& 20.0 & 2.9 & 4.7    & $>$6.2 & 1 \\ 
EES-Lynx 3  & 08$^h$48$^m$42$^s$.81 &+44\degg54\aminn34\asecc& 19.8 & 2.9 & 5.4    & $>$6.3 & 1 \\ 
EES-Lynx 4  & 08$^h$48$^m$48$^s$.93 &+44\degg57\aminn09\asecc& 18.9 & 2.4 & 5.4    & 6.5    & 1 \\
SA57 1      & 13$^h$09$^m$15$^s$.96 &+29\degg16\aminn50\asecc& 18.8 & 2.1 & 5.5    & 6.4    & 1 \\
SA57 2      & 13$^h$09$^m$19$^s$.21 &+29\degg20\aminn20\asecc& 19.4 & 2.5 & 5.3    & 6.3    & 1 \\
SA57 3      & 13$^h$09$^m$19$^s$.80 &+29\degg21\aminn05\asecc& 18.9 & 2.2 & 5.1    & 6.7    & 1 \\
EES-Pisces 1& 23$^h$12$^m$52$^s$.93 &+00\degg57\aminn06\asecc& 19.8 & 3.5 & 6.0    & $>$7.4 & 1 \\
EES-Pisces 2& 23$^h$12$^m$48$^s$.31 &+01\degg00\aminn09\asecc& 20.1 & 1.7 & 6.7    & $>$6.6 & 1 \\
EES-Cetus 1 & 03$^h$01$^m$26$^s$.85 &+00\degg21\aminn50\asecc& 19.9 & 2.3 & 5.1    & 6.5    & 1 \\
EES-Cetus 2 & 03$^h$01$^m$41$^s$.46 &+00\degg22\aminn32\asecc& 20.0 & 2.8 & 5.6    & 5.5    & 1 \\
EES-Cetus 3 & 03$^h$01$^m$24$^s$.74 &+00\degg22\aminn31\asecc& 20.2 & 2.3 & 5.8    & $>$6.0 & 1 \\
EES-Cetus 4 & 03$^h$01$^m$25$^s$.44 &+00\degg23\aminn18\asecc& 20.1 & 2.4 & $>$6.4 & $>$6.0 & 1 \\
EES-Cetus 5 & 03$^h$01$^m$24$^s$.97 &+00\degg25\aminn28\asecc& 20.0 & 2.2 & 5.4    & 6.2    & 1 \\
\hline
\multicolumn{8}{c}{\bf Other EROs from literature}\\
\hline
ERO-1       & 13$^h$12$^m$14$^s$.04 &+42\degg43\aminn55\asecc& 19.3 &...  & 6.2    &...     & 2 \\
ERO-2       & 16$^h$44$^m$57$^s$.08 &+46\degg26\aminn03\asecc& 18.7 & 2.7 & 6.2    &...     & 3 \\
ERO-3       & 17$^h$21$^m$46$^s$.14 &+50\degg02\aminn51\asecc& 19.6 &...  & 6.4    & 8.4    & 8 \\
ERO-5       & 21$^h$07$^m$15$^s$.29 &+23\degg31\aminn20\asecc& 19.7 &...  & $>$6.6 &...     & 4 \\
ERO-6       & 21$^h$07$^m$15$^s$.46 &+23\degg31\aminn38\asecc& 19.8 &...  & $>$6.5 &...     & 4 \\
ERO-7       & 22$^h$17$^m$33$^s$.24 &+00\degg16\aminn04\asecc& 19.6 &...  & $>$6.0 &...     & 5 \\
ERO-8       & 22$^h$17$^m$33$^s$.24 &+00\degg14\aminn28\asecc& 20.3 &...  & $>$6.0 &...     & 5 \\
J0905+3408  & 09$^h$05$^m$30$^s$.54 &+34\degg08\aminn09\asecc& 18.3 & ... & ...    & 6.1    & 6 \\
J1046$-$0016& 10$^h$46$^m$05$^s$.90 &$-$00\degg16\aminn46\asecc& 19.4 & ... & 5.1    & ...  & 7 \\
J1721+5002  & 17$^h$21$^m$44$^s$.50 &+50\degg02\aminn06\asecc& 20.0 & ... & 6.2    & 6.3    & 8 \\
J1019+0534  & 10$^h$19$^m$31$^s$.87 &+05\degg34\aminn36\asecc& 18.3 & ... & ...    & $>$7.0 & 9 \\
J1205$-$0743& 12$^h$05$^m$20$^s$.93 &$-$07\degg43\aminn51\asecc& 19.2 & 2.2 & 4.7    & 5.6  & 10 \\
J1249+3350  & 12$^h$49$^m$43$^s$.17 &+33\degg50\aminn04\asecc& 19.7 & ... & $>$4.9 & $>$6.3 & 11 \\
\hline\\
\multicolumn{8}{p{5in}}{$^a$ Though the sample was chosen with $K<\,\,$20, a revision of the
photometry after sample $~~~$selection resulted in a few objects with $K>\,\,$20.}\\
\multicolumn{8}{p{5in}}{1: Elston, R. et al. 2001, in preparation; 2: Cowie, L. L. et al. 1996, \aj, 112, 839; 3:
Hu, E. M. \& Ridgway, S. E. 1994, \aj, 107, 1303; 4: Knopp, G. P. \& Chambers, K. C. 1997,
\apjs, 109, 367; 5: Cowie, L. L. et al. 1994, \apj, 434, 114; 6: Eisenhardt, P.
\& Dickinson, M. 1992, \apj, 399, L47 ; 7: Moustakas, L. A. et al. 1997, \apj, 475, 445; 
8: McLeod, B. A. et al. 1995, \apjs, 96, 117; 9: Dey, A. et al. 1995, \apj, 440, 515; 
10: Giallongo, E. private communication; 11: Soifer, B. T. et al. 1994, \apj, 420, L1}\\
\end{tabular}
\end{center}
\end{table*}

\begin{table*}
\begin{center}
\caption[]{Log of VLA Observations}
\begin{tabular}{lclrcc}
\hline
\multicolumn{1}{c}{Field}&\multicolumn{1}{c}{\# of}&\multicolumn{1}{c}{Phase}&\multicolumn{1}{c}{Date of}&\multicolumn{1}{c}{rms}&\multicolumn{1}{c}{CLEAN}\\
\multicolumn{1}{c}{ }& \multicolumn{1}{c}{EROs}&\multicolumn{1}{c}{Calibrator}&\multicolumn{1}{c}{Obs.}&\multicolumn{1}{c}{($\mu$Jy)}&\multicolumn{1}{c}{beam}\\
\hline
\hline
\multicolumn{6}{c}{4.8 GHz VLA D Array Observations}\\
\hline
ERO-1      &  1 & 1327+434 & 28 May 1999 & 18 & 34\asecc$\times$17\asecc\\ 
ERO-5      &  1 & 2115+295 & 28 May 1999 & 15 & 20\asecc$\times$17\asecc\\ 
EES-Lynx   &  4 & 0818+423 & 28 May 1999 & 14 & 19\asecc$\times$16\asecc\\
SA57       &  3 & 1310+323 & 29 May 1999 & 14 & 27\asecc$\times$15\asecc\\
EES-Pisces &  2 & 2323$-$032 & 28 May 1999 & 16 & 25\asecc$\times$16\asecc\\
\hline
\multicolumn{6}{c}{1.4 GHz VLA B Array Observations}\\
\hline
EES-Pisces &  2 & 2323$-$032 & 12 Nov 1999 & 18 & 7\asecc$\times$5\asecc \\
EES-Cetus  &  5 & 0323+055 &  6 Nov 1999 & 75 & 6.5\asecc$\times$5.3\asecc \\
\hline
\end{tabular}
\end{center}
\end{table*}

\begin{table*}
\begin{center}
\caption[]{Radio, mm and sub-mm flux densities of EROs}
\begin{tabular}{lcccccccccc}
\hline
\multicolumn{1}{c}{Galaxy}& \multicolumn{1}{c}{1250$\mu$m} & \multicolumn{1}{c}{error} & \multicolumn{1}{c}{850$\mu$m} & \multicolumn{1}{c}{error} & \multicolumn{1}{c}{450$\mu$m} & \multicolumn{1}{c}{error} & \multicolumn{1}{c}{4.8 GHz} & \multicolumn{1}{c}{error} & \multicolumn{1}{c}{1.4 GHz} & \multicolumn{1}{c}{error} \\
\multicolumn{1}{c}{}& \multicolumn{1}{c}{(mJy)} & \multicolumn{1}{c}{(mJy)} & \multicolumn{1}{c}{(mJy)} & \multicolumn{1}{c}{(mJy)} & \multicolumn{1}{c}{(mJy)} & \multicolumn{1}{c}{(mJy)} & \multicolumn{1}{c}{($\mu$Jy)} & \multicolumn{1}{c}{($\mu$Jy)} & \multicolumn{1}{c}{($\mu$Jy)} & \multicolumn{1}{c}{($\mu$Jy)} \\
\hline
\hline
\multicolumn{11}{c}{\bf EES Sample}\\
\hline
EES-Lynx 1   & ...    & ...  & $-$1.6 & 1.8 &... & ...& ~19.0   & 14.0 & ...     & ...   \\
EES-Lynx 2   & ...    & ...  & $-$0.1 & 2.0 &... & ...& ~10.6   & 14.0 & ...     & ...   \\ 
EES-Lynx 3   & ...    & ...  & ~~1.2  & 1.7 &... & ...& ~~~5.4  & 14.0 & ...     & ...   \\ 
EES-Lynx 4   & $-$0.1 & 0.3  & $-$0.8 & 2.6 &... & ...& $~-$4.9 & 14.0 & ...     & ...   \\
SA57 1       & ~~{\bf 1.5$^a$} & {\bf 0.5} & ~~2.9 & 1.6 &...& ...& $-$12.0 & 14.0 & ...& ...\\
SA57 2       & ~~0.8  & 0.4  & ~~3.5  & 2.3 &... & ...& $~-$5.0 & 14.0 & ...     & ...   \\
SA57 3       & $-$0.3 & 0.4  & $-$5.6 & 2.7 &... & ...& ~~16.0  & 14.0 & ...     & ...   \\
EES-Pisces 1 & ...    & ...  & $-$1.9 & 1.9 &... & ...& ~~~2.6  & 16.0 & $-$11.0 & 18.0  \\
EES-Pisces 2 & $-$0.3 & 0.2  & $-$1.0 & 1.9 &... & ...& $~-$1.5 & 16.0 & $-$27.0 & 18.0  \\
EES-Cetus 1  & ...    & ...  & ~~0.7  & 2.4 & 11 & 45 &... & ...& ~{\bf 520.0$^a$}&{\bf 75.0}\\
EES-Cetus 2  & ~~0.2  & 0.4  & ~~2.4  & 2.8 & 128& 118& ... &...& ~{\bf 380.0$^a$}&{\bf 75.0}\\
EES-Cetus 3  & ...    & ...  & $-$0.9 & 2.1 & 5  & 29 & ...     &  ... & $-$14.0 & 75.0  \\
EES-Cetus 4  & ~~0.5  & 0.4  & ~~1.9  & 2.1 & 17 & 23 & ...     &  ... & ~~70.0  & 75.0  \\
EES-Cetus 5  & ...    & ...  & ~~1.8  & 2.3 & 33 & 32 & ...     &  ... & ~~47.0  & 75.0  \\
\hline
\multicolumn{11}{c}{\bf Other EROs from literature}\\
\hline
ERO-1        & ~~0.7  & 1.4  & ~~1.7 & 1.1 & ... &... & ~~0.4 & 18.0 & $<$14$^{b}$  & ... \\
ERO-2        & $-$0.8 & 1.3  & ~~1.1 & 4.7 & ... &... &...    & ...  & ...  & ... \\
ERO-3        & ~~0.1  & 1.1  & $-$0.9& 3.9 & ... &... & ...   & ...  & ...  & ... \\  
ERO-5        &...     & ...  & ~~2.2 & 1.1 & ... &... & $<$70$^c$ & ...  & ...  & ... \\
ERO-6        &...     & ...  & ~~3.1 & 2.9 & ... &... & ...   & ...  & ...  & ... \\    
ERO-7        &...     & ...  & ~~4.2 & 2.2 & ... &... & ...   & ...  & ...  & ... \\ 
ERO-8        &...     & ...  & $-$0.5& 2.1 & ... &... & ...   & ...  & ...  & ... \\ 
J0905+3408   & ~~0.1  & 0.9  & ...   & ... & ... &... & ...   & ...  & ...  & ... \\
J1046$-$0016 & ~~0.4  & 1.7  & ...   & ... & ... &... & ...   & ...  & ...  & ... \\
J1721+5002   & ~~0.4  & 1.1  & ...   & ... & ... &... & ...   & ...  & ...  & ... \\
J1019+0534   & $-$1.6 & 1.7  & ...   & ... & ... &... & ...   & ...  & ...  & ... \\
J1205$-$0743 & $-$0.7 & 1.7  & ...   & ... & ... &... & ...   & ...  & ...  & ... \\
J1249+3350   & $-$0.7 & 1.3  & ...   & ... & ... &... & ...   & ...  & ...  & ... \\
\hline
\multicolumn{11}{p{6in}}{$^a$ The three entries in bold face are detections. 
The rest are the 
flux densities measured at the NIR positions of the EROs, with the associated 
1$\sigma$ error. These values have been used to calculate the statistical flux 
densities of the sample (see Sect. 4).}\\
\multicolumn{11}{p{6in}}{$^b$ 3$\sigma$ upper limit : Richards, private communication.}\\
\multicolumn{11}{p{6in}}{$^c$ The upper limit to the 4.8 GHz flux density of ERO-5 is 
quoted as this source is 1.5 synthesized beams away from a strong confusing source.}\\
\end{tabular}
\end{center}
\end{table*}

\end{document}